\documentclass[aps, superscriptaddress]{revtex4-1}
\pdfoutput=1
\usepackage{graphicx}

\usepackage{latexsym}
\usepackage[fleqn]{amsmath}
\usepackage{bm}
\usepackage{color}
\usepackage{amssymb}
\usepackage{bm}
\usepackage{ascmac}
\usepackage{bm}
\usepackage{comment}

\newcommand{\la}{\left\langle}
\newcommand{\ra}{\right\rangle}

\newcommand{\abs}[1]{{|#1|}}

\begin{document}

	\title{Optimization of periodic input waveforms for global entrainment of weakly forced limit-cycle oscillators}
	\author{Yuzuru Kato}
	\email{Corresponding author: kato.y.bg@m.titech.ac.jp}
	\affiliation{Department of Systems and Control Engineering,
		Tokyo Institute of Technology, Tokyo 152-8552, Japan}
	
	\author{Anatoly Zlotnik}
	\affiliation{Applied Mathematics and Plasma Physics, Los
		Alamos National Laboratory, Los Alamos, NM 87545}
	
	\author{Jr-Shin Li}
	\affiliation{Department of Electrical and Systems Engineering, Washington University in St. Louis, St. Louis, Missouri 63130, USA}
	
	\author{Hiroya Nakao}
	\affiliation{Department of Systems and Control Engineering,
	Tokyo Institute of Technology, Tokyo 152-8552, Japan}
	\date{\today}
	
\begin{abstract}
We propose a general method for optimizing periodic input waveforms for global entrainment of weakly forced limit-cycle oscillators based on phase reduction and nonlinear programming. We derive averaged phase dynamics from the mathematical model of a limit-cycle oscillator driven by a weak periodic input and optimize the Fourier coefficients of the input waveform to maximize prescribed objective functions. In contrast to the optimization methods that rely on the calculus of variations, the proposed method can be applied to a wider class of optimization problems including global entrainment objectives. As an illustration, we consider two optimization problems, one for achieving fast global convergence of the oscillator to the entrained state and the other for realizing prescribed global phase distributions in a population of identical uncoupled noisy oscillators. We show that the proposed method can successfully yield optimal input waveforms to realize the desired states in both cases.
\end{abstract}
\maketitle

\section{Introduction}
Rhythms and synchronization are 
observed widely in various fields of science and technology, 
such as electrical oscillations, chemical oscillations, biological 
rhythms, and mechanical vibrations \cite{winfree2001geometry, kuramoto1984chemical, 
ermentrout2010mathematical, pikovsky2001synchronization, glass1988clocks, strogatz1994nonlinear}.
Engineering  applications of synchronization 
have also been considered, such as 
injection locking \cite{adler1946study, kurokawa1973injection} and
phase lock loops \cite{best1984phase} in electrical circuits, 
Josephson voltage standard~\cite{josephson1962possible, shapiro1963josephson}, and deep brain stimulation for the treatment of Parkinson's disease and epilepsy seizure \cite{tass2001desynchronizing, good2009control}.

Rhythmic dynamical systems are modeled typically as nonlinear limit-cycle oscillators.
Under the assumption that the perturbation applied to the limit-cycle oscillator is 
sufficiently weak,
we can approximately derive a one-dimensional phase equation 
describing the oscillator dynamics by using the phase reduction theory~\cite{winfree2001geometry, kuramoto1984chemical,  ermentrout2010mathematical, nakao2016phase, ermentrout1996type,brown2004phase}. 
The simplicity of the phase equation has facilitated systematic analysis of the universal dynamics
of limit-cycle oscillators, such as entrainment of oscillators by periodic forcing and mutual synchronization between coupled oscillators.
The collective synchronization transition in a large population of coupled phase oscillators~\cite{kuramoto1984chemical} is one of the most well-known results predicted by using the phase equation. 
The phase reduction theory has also been extended to non-conventional physical systems, such as piecewise-smooth oscillators~\cite{shirasaka2017phase2}, rhythmic spatiotemporal patterns~\cite{kawamura2013collective,nakao2014phase}, 
and quantum limit-cycle oscillators \cite{kato2019semiclassical}.

Using the phase equation, we can also formulate 
optimization and control of 
nonlinear oscillators \cite{monga2019phase},
for example, minimizing the power for control of oscillators~\cite{moehlis2006optimal, 
dasanayake2011optimal, zlotnik2012optimal, li2013control},
maximizing the range~\cite{harada2010optimal, tanaka2014optimal, tanaka2015optimal, yabe2020locking}
or linear stability~\cite{zlotnik2013optimal} of entrainment for periodically forced oscillators,
maximizing linear stability of mutual synchronization between two coupled oscillators~\cite{shirasaka2017optimizing, watanabe2019optimization}, 
maximizing phase coherence of noisy oscillators~\cite{pikovsky2015maximizing},
phase-selective entrainment of oscillators~\cite{zlotnik2016phase},
control of phase distributions in oscillator populations~\cite{monga2018synchronizing, kuritz2019ensemble, monga2019phase2}, and optimizing 
entrainment stability of quantum limit-cycle oscillators in the semiclassical regime \cite{kato2020semiclassical}.

In previous studies, optimization problems 
for the improvement of entrainment range,
entrainment stability, and enhancement of phase coherence have been 
formulated mainly using the calculus of variations
\cite{harada2010optimal, zlotnik2013optimal, pikovsky2015maximizing, moehlis2006optimal, dasanayake2011optimal}.
This formulation often 
gives explicit analytical 
solutions of the optimization problems and provides us with theoretical insights.
For example, the analogy between the optimal waveform and the proportional-integral-differential (PID) controller in the feedback control theory \cite{aastrom1995pid} has been discussed \cite{kato2020semiclassical}.
However, for more general oscillator dynamics, the optimization problems may not be solvable by using the calculus of variations.

\begin{figure} [!t]
	\begin{center}
		\includegraphics[width=1.0\hsize,clip]{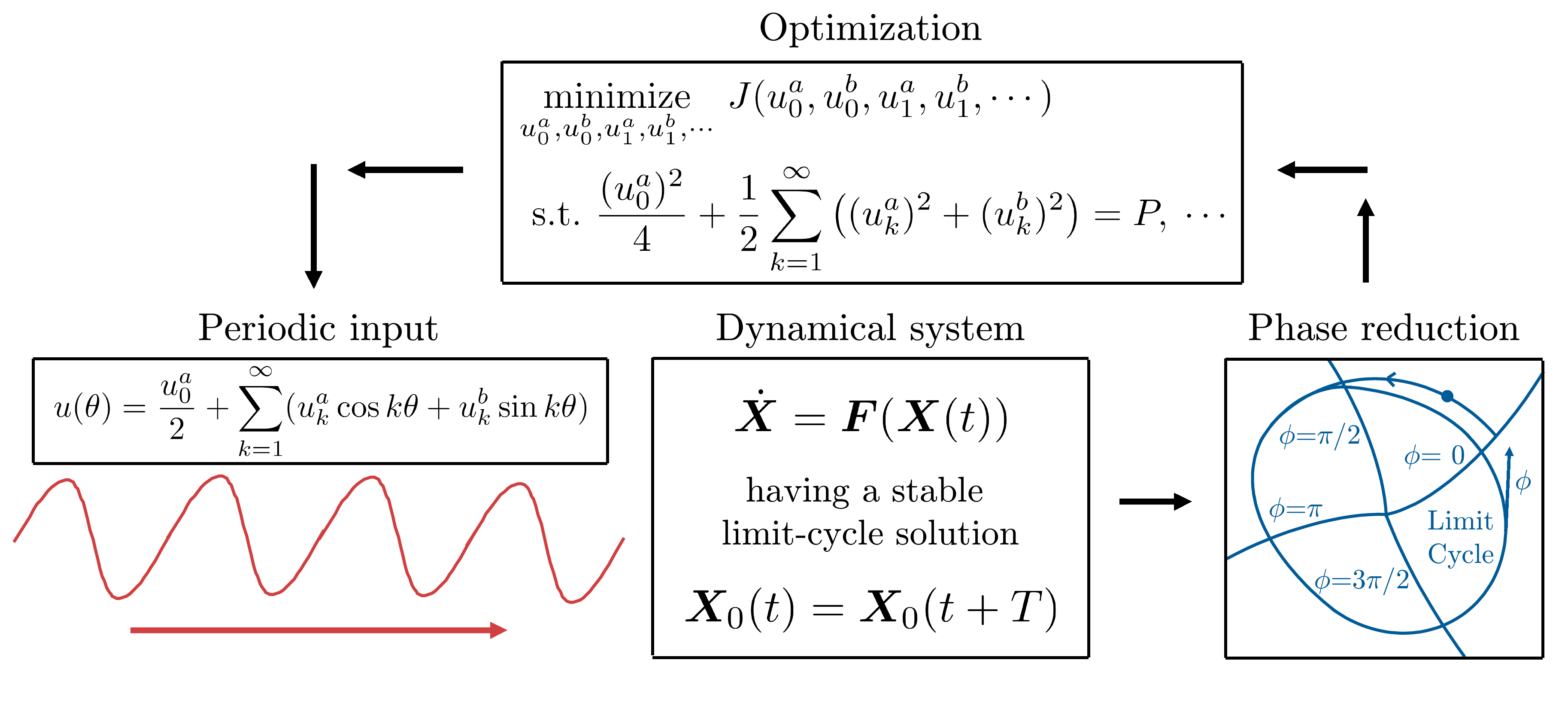}
		\caption{
			A schematic diagram showing 
			the optimization framework for entrainment of 
			weakly forced limit-cycle oscillators in which Fourier 
			coefficients of the periodic input waveforms are numerically 
			optimized by using nonlinear programming. 
		}
		\label{fig_1}
	\end{center}
\end{figure}

In this paper,
we propose a method for optimizing periodic input waveforms for global entrainment of weakly forced limit-cycle oscillators based on phase reduction and nonlinear programming,
which can be applied to a wide range of optimization problems 
that cannot be solved analytically by using the calculus of variations.
In this method, the desired oscillator dynamics are represented by Fourier coefficients of the periodic
input waveforms and these coefficients are numerically optimized.
A schematic diagram of the proposed method is shown in Fig.~\ref{fig_1}.
In contrast to the preceding studies that developed a Lyapunov-based control framework 
\cite{monga2018synchronizing, kuritz2019ensemble}
and an optimal control framework by Monga and Moehlis~\cite{monga2019phase2} for the non-averaged phase equation, our present framework is based on the averaged phase equation and thus gives a simpler method for finding the optimal waveforms and can easily incorporate the effect of weak noise.
We solve two optimization problems to illustrate the proposed method,
both of which are not analytically solvable by using the calculus of variations.

\section{Theory}

\subsection{Periodically forced limit-cycle oscillator}

Various rhythmic systems in the real world can be modeled as limit-cycle oscillators. Here, the limit-cycle oscillator stands for a nonlinear dynamical system with a stable and isolated periodic orbit, which can exhibit stable self-sustained oscillations even under the effect of weak perturbations~\cite{strogatz1994nonlinear}.
In this study, we consider 
a limit-cycle oscillator subjected to a weak periodic forcing.
Assuming that the first variable of the system receives the periodic forcing input, our model is given by
\begin{align}
\label{eq:X}
\dot{\bm{X}}(t) = {\bm{F}}({\bm X}(t)) + 
u(\omega_e t)(1, 0, \ldots, 0)^\mathsf{T}, 
\end{align}
where $\bm X \in {\mathbb R}^{N \times 1}$ represents a system state, 
$\bm{F}({\bm X}) \in {\mathbb R}^{N \times 1}$ is a smooth vector field,
$u(\omega_e t)$ is a $2\pi$-periodic smooth scalar function representing 
the periodic input force with frequency $\omega_e$, and $(1, 0, \ldots, 0)^\mathsf{T} \in {\mathbb R}^{N \times 1}$ is a vector where $\mathsf{T}$  represents the transpose.
We also assume that the amplitude of the forcing input $u$ is sufficiently small and of $O(\varepsilon)$ where $0 < \varepsilon \ll 1$. 

We assume that 
in the absence of a periodic input, the system $\dot{{\bm{X}}} = {\bm{F}}({\bm{X}})$
has an exponentially stable limit-cycle solution ${\bm{X}}_{0}(t) = {\bm{X}}_{0}(t+T)$ 
with a natural period $T$ and frequency $\omega = 2\pi / T$. 
Following the standard method of phase reduction~\cite{winfree2001geometry, kuramoto1984chemical,  ermentrout2010mathematical, nakao2016phase, ermentrout1996type,brown2004phase}, we can  introduce an asymptotic phase function $\Phi({\bm{X}}) : {\mathbb R}^{N \times 1} \to [0, 2\pi)$ such that 
$\nabla \Phi({\bm{X}}) \cdot {\bm F}({\bm{X}})  = \omega$
is satisfied in the basin of the limit cycle, where $\nabla \Phi({\bm X}) \in {\mathbb R}^{N \times 1}$ is the gradient of $\Phi({\bm X})$ ($\cdot{}$ denotes a scalar product of two vectors).
The phase of a system state ${\bm X}$ is defined as $\phi = \Phi({\bm X})$, which satisfies $\dot{\phi} = \dot{\Phi}({\bm X}) = {\bm F}({\bm X}) \cdot \nabla \Phi({\bm X}) = \omega$.
We represent the system state ${\bm X}$ on the limit cycle as ${\bm X}_0(\phi)$ as a function of the phase $\phi$. Note that an identity $\Phi({\bm X}_0(\phi)) = \phi$ is satisfied 
by the definition of $\Phi({\bm X})$.

Using the chain rule of differentiation, we can derive an equation for the phase $\phi$ 
under the effect of the periodic input as
$\dot{\phi} = \omega +  \nabla \Phi({\bm X}) \cdot u(\omega_e t)(1, 0, \ldots, 0)^\mathsf{T}$, which is still not closed in $\phi$ because $\nabla \Phi({\bm X})$ depends on ${\bm X}(t)$.
Because we assume that the periodic input is sufficiently weak and 
the deviation of the state ${\bm X}(t)$ from the limit cycle is small, 
at the lowest-order approximation, we can approximate ${\bm X}(t)$ by ${\bm X}_0(\phi(t))$ and derive the equation for the phase $\phi$ as
\begin{align}
\label{eq:dphi}
\dot{\phi} &= \omega + \bm{Z}( \phi ) \cdot u(\omega_e t)(1, 0, \ldots, 0)^\mathsf{T},
\end{align}
which is correct up to $O(\varepsilon)$. We here introduce the phase sensitivity function (PSF) $\bm{Z}(\phi) = \nabla \Phi|_{ {\bm{X} }  = {\bm{X}}_{0}(\phi)} \in {\mathbb R}^{N \times 1}$ that characterizes
the linear response of the oscillator phase to weak perturbations.
This PSF can be numerically obtained as a $2 \pi$-periodic solution to an adjoint-type equation of Eq.~(\ref{eq:X}) with appropriate normalization \cite{ermentrout2010mathematical}. 

To formulate the optimization problem, we further derive an averaged phase equation from the phase equation (\ref{eq:dphi}).
We assume that the frequency $\omega_e$ of the periodic input is close to the natural frequency $\omega$ of the limit cycle
and introduce a phase difference $\psi = \phi - \omega_e t$ between the oscillator and periodic modulation, which is a slow variable obeying 
\begin{align}
\label{eq_psi}
\dot{\psi} &=   \Delta_e + Z_1( \psi + \omega_e t) u(\omega_e t),
\end{align}
where $\Delta_e = \omega - \omega_e$ is assumed to be $O(\varepsilon)$
and
$Z_1$ is the first component of the PSF.
Following the standard averaging procedure~\cite{kuramoto1984chemical}, the small right-hand side of this equation can be averaged over one-period of oscillation of the unperturbed system  (note that both $\Delta_{e}$ and $u$ are $O(\varepsilon)$), yielding an averaged phase equation,
\begin{align}
\label{eq:dpsi}
\dot{\psi} =  \left\{ \Delta_e + \Gamma(\psi) \right\},
\end{align}
which is correct up to $O(\varepsilon)$. Here, $\Gamma(\psi)$ is a smooth $2\pi$-periodic  phase coupling function defined as
\begin{align}
\label{eq:gamma}
\Gamma(\psi) 
= \la Z_1(\psi + \theta) u(\theta) \ra_\theta,
\end{align}
where the one-period average is denoted as 
$\la \cdot \ra_{\theta} = \frac{1}{2\pi} \int_0^{2\pi} (\cdot) d\theta$.
If the dynamics of $\psi$ has a stable fixed point at $\psi_s$, the oscillator state can be entrained to the periodic input with the phase difference $\psi_s$. 
By varying the functional form of $\Gamma(\psi)$, the dynamics of the oscillator phase can be controlled.

\subsection{Optimization problem and Fourier representation}

Our aim is to realize desired dynamics of the oscillator (or population of uncoupled oscillators) by optimizing the waveform $u$ of the input to yield an appropriate phase coupling function $\Gamma$ under a constraint on the input power.
Introducing an objective function $\tilde{J}(\Gamma)$ as a function of $\Gamma$
and assuming that the input power is fixed as $\langle u^2(\theta) \rangle_\theta = P$ where $P$ is of $O(\varepsilon^2)$, we can formulate the following optimization problem:
\begin{align}
\label{eq:opt_obj}
&\underset{u}{\mbox{minimize}}\ \tilde{J}(\Gamma)
\quad \mbox{s.t.} \quad
\langle u^2(\theta) \rangle_\theta = P
\ \mbox{and} \ 
[\cdots],
\end{align}
where $[\cdots]$ represents additional constraints that can also be included in the formulation.

To solve this problem, we use the Fourier series to write the PSF and periodic input as
\begin{align}
\label{eq:fourier}
Z_1(\theta) &= \frac{z^a_{0}}{2} + \sum_{k = 1}^{\infty} ( z^a_{k} \cos k\theta + z^b_{k} \sin k \theta ), 
\cr
u(\theta) &= \frac{u^a_{0}}{2} + \sum_{k = 1}^{\infty} ( u^a_{k} \cos k \theta + u^b_{k} \sin k \theta ), 
\end{align}
where
\begin{align}
z^a_k &= 
\frac{1}{\pi} \int_{0}^{2\pi}  Z_1(\theta) \cos k \theta d\theta,~ 
z^b_k = \frac{1}{\pi} \int_{0}^{2\pi}  Z_1(\theta) \sin k \theta  d\theta, 
\cr
u^a_k &= 
\frac{1}{\pi} \int_{0}^{2\pi}  u(\theta) \cos k \theta d\theta,~
u^b_k = \frac{1}{\pi} \int_{0}^{2\pi}  u(\theta) \sin k \theta d\theta.
\end{align}
The phase coupling function can then be written as
\begin{align}
\label{eq:gamma2}
{\Gamma}(\psi) = 
\frac{1}{4} z^a_0 u^a_0 + 
\frac{1}{2} \sum_{k = 1}^{\infty}
\left[  (z^a_k \cos k\psi + z^b_k \sin k\psi)u^a_k +  (z^b_k \cos k\psi - z^a_k \sin k\psi)u^b_k \right].
\end{align}

Considering that $\Gamma(\psi)$ is parametrized 
by the Fourier coefficients $u^a_k, u^b_k~(k =0,1,2,\ldots)$, where $u^b_0 = 0$, 
the objective function $\tilde{J}(\Gamma)$ can be regarded as a function of the Fourier coefficients, namely, 
$\tilde{J}(\Gamma) = J(u^a_0, u^b_0, u^a_1, u^b_1, \cdots)$,
and we can rewrite the optimization problem (\ref{eq:opt_obj}) as
\begin{align}
\label{eq:opt_obj2}
&\underset{u^a_0, u^b_0, u^a_1, u^b_1, \cdots}{\mbox{minimize}}\ 
J(u^a_0, u^b_0, u^a_1, u^b_1, \cdots)
\quad \mbox{s.t.} \quad
\frac{(u^{a}_0)^2}{4} + \frac{1}{2} \sum_{k = 1}^{\infty} \left( (u^{a}_k)^2 +  (u^{b}_k)^2 \right) = P
\ \mbox{and}\ 
[\cdots].
\end{align}
The optimal waveform of the periodic input can be calculated from Eq.~(\ref{eq:fourier})
using the Fourier coefficients $u^a_k$ and $u^b_k~(k =0,1,2,\ldots)$ obtained by solving
the optimization problem (\ref{eq:opt_obj2}).

In practice, the Fourier coefficients $z_k^a$ and $z_k^b$ of the PSF decay quickly with the wavenumber $k$, so the summation in $\Gamma(\psi)$ can be truncated at some finite number $k_{max}$ by neglecting small-amplitude Fourier coefficients satisfying $\sqrt{ (z^a_k)^2  + (z^b_k)^2} = \delta$ with a small parameter $\delta$  ($0 < \delta \ll 1$).
Thus, we only need to optimize a finite number $k_{max}$ of the Fourier coefficients $u^a_k$ and $u^b_k$ of the periodic input waveform.
The fact that the PSF has only a limited number of harmonics in practice also imposes a fundamental limitation to the realizability of the oscillator dynamics as we discuss later.
We set $\delta = 0.001$ in this study.

We here note that the derived optimal waveform is given to the system in the form of $u(\theta = \omega_e t)$ in Eq.~(\ref{eq:X}). Therefore, the proposed method provides a feedforward control that can be implemented without measuring the system.
Also, the present formulation is applicable to a wider class of optimization problems than those previously formulated by using  the averaging method and calculus of variations.
In Appendix A,  by using the present formulation, we derive the  results of Refs.~\cite{harada2010optimal, zlotnik2013optimal, pikovsky2015maximizing}  
previously obtained by  the averaging method and calculus of variations. 
For realistic oscillators (including the FitzHugh-Nagumo oscillator with the parameter set used in this study), 
the number $k_{max}$ is typically smaller than $20$
and we can easily solve the optimization problem using appropriate numerical solvers for nonlinear programming.
In this study, we use the scipy.optimize.minimize toolbox with the 
 sequential least squares programming (SLSQP) method for Python to solve the optimization problems, which minimizes a scalar function of one or more variables~\cite{virtanen2020scipy}. 

\section{Achieving fast global entrainment}
\label{optimization1}
First, we apply the proposed method for achieving fast 
global convergence of the oscillator to the entrained state.
This kind of problem is important because the demand for quick adjustment of the oscillator rhythms arises in various situations, for example, achieving rapid cardiac resynchronization \cite{ermentrout1984beyond} and 
quick recovery from jet lag \cite{waterhouse2007jet}.
In a previous study \cite{zlotnik2013optimal}, the optimization problem for the 
local convergence property characterized by the linear stability
of the entrained state has been solved by using the calculus of variations.
However, as we show below in the examples, maximization of the local linear stability property can sometimes deteriorate the global convergence property.
Here, rather than maximizing the local linear stability, we minimize the average convergence time to the entrained state in order to realize fast global entrainment from wide initial conditions.

\subsection{Formulation of the optimization problem}

We assume that Eq.~(\ref{eq:dpsi}) has only a single pair of fixed points, namely, a single stable fixed point at $\psi=0$ (we can assume this without loss of generality by shifting the origin of the oscillator phase) satisfying $\Delta_e + \Gamma(0) = 0$ and $\Gamma'(0) < 0$, 
and a single unstable fixed point at $\psi = \psi^{*}$ satisfying $\Delta_e + \Gamma(\psi^{*}) = 0$ with $\Gamma'(\psi^{*}) > 0$.
The stable fixed point at $0$ corresponds to the entrained state of the oscillator to the periodic input.
The location $\psi^{*}$ of the unstable fixed point is determined by 
the functional form of 
$\Gamma$ and we denote it as $\psi_{\Gamma}^{*}$ in the following discussion.

We use small parameters $\epsilon_f$ and $\epsilon_c$  ($0 < \epsilon_f, \epsilon_c \ll 1$) to characterize the neighborhoods of the fixed points.
We define an $\epsilon_f$-neighborhood of the stable fixed point at $0$ as 
$\mathcal{B}=\{\varphi \in [-\pi, \pi] ~|~ \abs{\varphi } < \epsilon_f\}$  and an $\epsilon_c$-neighborhood~($0 < \epsilon_c \ll 1$) of the unstable fixed point at $\psi_{\Gamma}^{*}$ as  
$\mathcal{C}_{\Gamma}=\{\varphi \in [-\pi, \pi]  ~|~ \abs{\varphi - \psi_{\Gamma}^{*}} < \epsilon_c\}$,  where the subscript $\Gamma$ of $\mathcal{C}$ represents the dependence of $\mathcal{C}$ on $\Gamma$ through the location of the unstable fixed point $\psi_{\Gamma}^{*}$.
Here, the phase values and their differences are considered in modulo $2\pi$.
For simplicity, we consider the case that the stable and unstable fixed points are not too close to each other and assume $\mathcal{B} \cap {\mathcal C}_{\Gamma} =  \emptyset$. 
This also ensures that the proposed method will work even if small approximation error is caused by the phase reduction.
We define the set of all points in $[-\pi, \pi]$ except for those in the neighborhood $\mathcal{B}$ or ${\mathcal C}_{\Gamma}$ of either of the fixed points $0$ and $\psi_{\Gamma}^{*}$ as 
${\mathcal A}_{\Gamma} = [ -\pi, \pi ] \setminus ( \mathcal{B} \cup {\mathcal C_{\Gamma}} ) = \{ \varphi \in [-\pi, \pi] ~|~ \varphi \notin \mathcal{B},\ {\mathcal C}_{\Gamma} \}$.
It is noted that the set ${\mathcal A}_{\Gamma}$ also depends on $\Gamma$ through $\psi_{\Gamma}^{*}$.

We define $T_{\Gamma}(\psi_0 \to \mathcal{B})$ as the convergence time 
from the initial point 
$\psi = \psi_0 \in \mathcal{A}_{\Gamma}$
to the entrained state
$\psi \in \mathcal{B}$, which is determined by the functional form of $\Gamma$.
In the limit $\epsilon_f \to 0$, the convergence time $T_{\Gamma}(\psi_0 \to \mathcal{B})$ is dominated by the dynamics near the fixed points and diverges to infinity.
Also, in the limit $\epsilon_c \to 0$, the initial phase difference can be arbitrarily close to the unstable fixed point $\psi_{\Gamma}^{*}$ and the convergence time from such a point to $\mathcal{B}$ tends to diverge logarithmically.
We do not consider these limits but rather keep $\epsilon_f$ and $\epsilon_c$ small but finite so that the convergence time $T_{\Gamma}(\psi_0 \to \mathcal{B})$ 
characterizes the time necessary for the oscillator phase to enter the entrained region within the allowed precision.  
Here, though mathematically the logarithmic singularity around $\psi_{\Gamma}^{*}$ is integrable, we also exclude this logarithmic divergence because it can dominate the convergence time numerically and be unfavorable for the optimization.
\begin{figure} [!t]
	\begin{center}
		\includegraphics[width=1.0\hsize,clip]{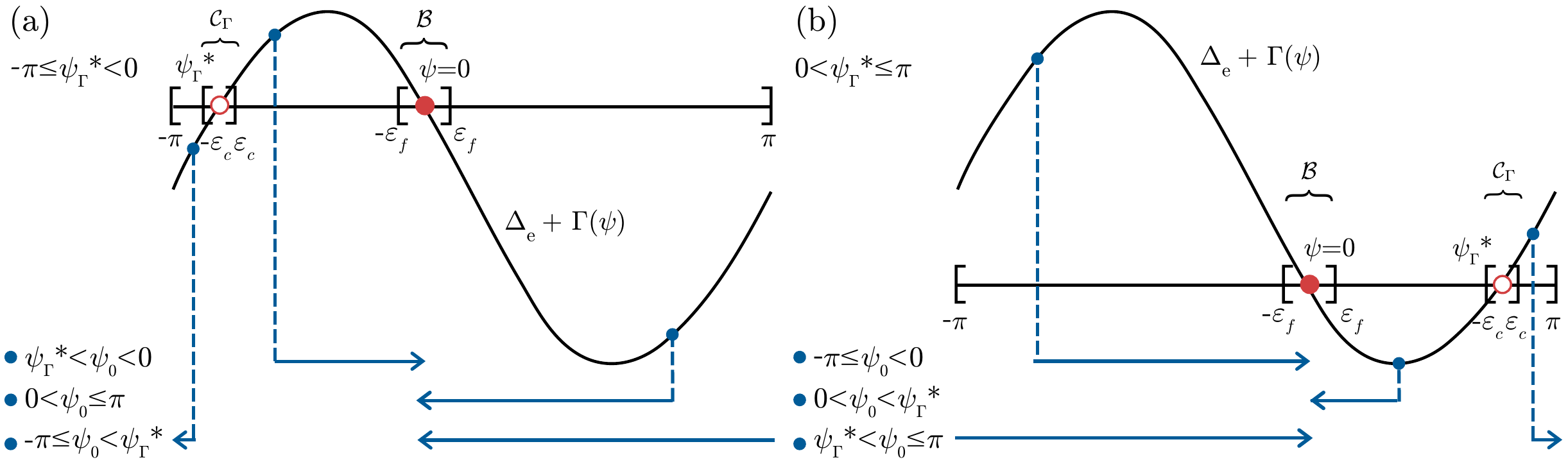}
		\caption{
			Schematic diagrams showing convergence of the phase difference $\psi$ from the initial value $\psi_0$ to the entrained state ${\mathcal B}$. Depending on {the location of the unstable fixed point} $\psi^*_{\Gamma}$ and {the initial condition} $\psi_0$, $\psi$ approaches ${\mathcal B}$ from either the positive or negative side as depicted by the arrows.
			 $(a)~-\pi \leq \psi_{\Gamma}^{*} < 0$,\ \ $(b)~ 0 < \psi_{\Gamma}^{*} \leq \pi$.
		}
		\label{fig_2}
	\end{center}
\end{figure}

Depending on the location of the unstable fixed point $\psi^*_{\Gamma}$ and the initial value $\psi_0$ of the phase difference $\psi$, the convergence time $T_{\Gamma}(\psi_0 \to \mathcal{B})$ 
of $\psi$ to $\mathcal{B}$ is expressed in 6 ways as follows. See Fig.~\ref{fig_2} for schematic diagrams.

First, when $-\pi \leq \psi_{\Gamma}^{*} < 0$ (Fig.~\ref{fig_2}(a)),
\begin{align}
T_{\Gamma}(\psi_0 \to \mathcal{B}) =
\left\{
\begin{array}{l}
	\int_{\psi_0}^{-\pi} ( \Delta_e + \Gamma(\theta) )^{-1} d\theta + \int_{\pi}^{\epsilon_f}  ( \Delta_e + \Gamma(\theta) )^{-1} d\theta \quad ( -\pi \leq \psi_0 < \psi_{\Gamma}^{*} ),
\cr
\cr
	\int_{\psi_0}^{- \epsilon_f} ( \Delta_e + \Gamma(\theta) )^{-1} d\theta \quad ( \psi_{\Gamma}^{*} < \psi_0 < 0),
\cr
\cr
	\int_{\psi_0}^{+ \epsilon_f} ( \Delta_e + \Gamma(\theta) )^{-1} d\theta \quad ( 0 < \psi_0 \leq \pi ).
\end{array}
\right.
\end{align}
Note that in the second case ($\psi_{\Gamma}^{*} < \psi_0 < 0$), 
$\Delta_e + \Gamma(\theta) > 0$
and $\psi$ approaches $\mathcal{B}$ from the negative side, and in the first and third cases, 
$\Delta_e + \Gamma(\theta) < 0$
and $\psi$ approaches $\mathcal{B}$ from the positive side.
Similarly, when $0 < \psi_{\Gamma}^{*} \leq \pi$ (Fig.~\ref{fig_2}(b)),
\begin{align}
T_{\Gamma}(\psi_0 \to \mathcal{B}) =
\left\{
\begin{array}{l}
	\int_{\psi_0}^{-\epsilon_f} ( \Delta_e + \Gamma(\theta) )^{-1} d\theta \quad ( -\pi \leq \psi_0 < 0 ),
\cr
\cr
	\int_{\psi_0}^{+\epsilon_f} ( \Delta_e + \Gamma(\theta) )^{-1} d\theta \quad ( 0 < \psi_0 < \psi_{\Gamma}^{*}),
\cr
\cr
	\int_{\psi_0}^{\pi} ( \Delta_e + \Gamma(\theta) )^{-1} d\theta + \int_{-\pi}^{- \epsilon_f} ( \Delta_e + \Gamma(\theta) )^{-1} d\theta \quad ( \psi_{\Gamma}^{*} < \psi_0 \leq \pi ).
\end{array}
\right.
\end{align}
In the second case ($0 < \psi_0 < \psi_{\Gamma}^{*}$), 
 $\Delta_e + \Gamma(\theta) < 0$
and $\psi$ approaches $\mathcal{B}$ from the positive side (and from the negative side in the other two cases).
In the above expressions, $(\Delta_e + \Gamma(\theta))^{-1} d\theta$ gives the time necessary for the phase difference to change from $\theta$ to $\theta+d\theta$. 

To characterize the speed of global convergence of the oscillator to the entrained state given the functional form $\Gamma$, we consider the average convergence time of the phase difference from initial points in ${\mathcal A}_{\Gamma}$ to the $\epsilon_f$-neighborhood $\mathcal{B}$ of the stable fixed point $0$, defined as
\begin{align}
T_{ave}(\Gamma) = 
\frac{1}{| \mathcal{A}_\Gamma |} \int_{\psi_0 \in \mathcal{A}_{\Gamma}} T_{\Gamma}(\psi_0 \to \mathcal{B}) d \psi_0 
\end{align}
where $| \mathcal{A}_\Gamma | = 2(\pi - \epsilon_f - \epsilon_c) $ is the size of $\mathcal{A}_\Gamma$.
Here, we take the average over uniformly distributed initial points $\psi_0$ in the set $\mathcal{A}_{\Gamma}$.
To realize fast global convergence of the oscillator to the entrained state, we minimize this average time $T_{ave}(\Gamma)$ by optimizing $\Gamma$. The optimization problem, Eq.~(\ref{eq:opt_obj2}), is formulated in this case as follows:
\begin{align}
\label{eq:opt_obj_gcs}
\underset{u^a_0, u^b_0, u^a_1, u^b_1, \cdots}{\mbox{minimize}}\ 
&
\frac{1}{| \mathcal{A}_\Gamma |} \int_{\psi_0 \in \mathcal{A}_{\Gamma}}
T_{\Gamma}(\psi_0 \to \mathcal{B}) d\psi_0
\cr
\quad \mbox{s.t.} \quad  
&
\frac{(u^{a}_0)^2}{4} + \frac{1}{2} \sum_{k = 1}^{\infty} \left( (u^{a}_k)^2 +  (u^{b}_k)^2 \right) = P,
\cr
&
\Delta_e + {\Gamma}(0) = 0,
\quad
{\Gamma}'(0) < 0,
\cr
&
{\mathcal B} \cap {\mathcal C}_{\Gamma} = \emptyset.
\end{align}
Here, the additional constraints represent that $\psi=0$ is a linearly stable fixed point and that the stable and unstable fixed points are not too close.

 %
 %
 \subsection{FitzHugh-Nagumo model}

 \label{opt1_model}
 \begin{figure} [!t]
 	\begin{center}
 		\includegraphics[width=1.0\hsize,clip]{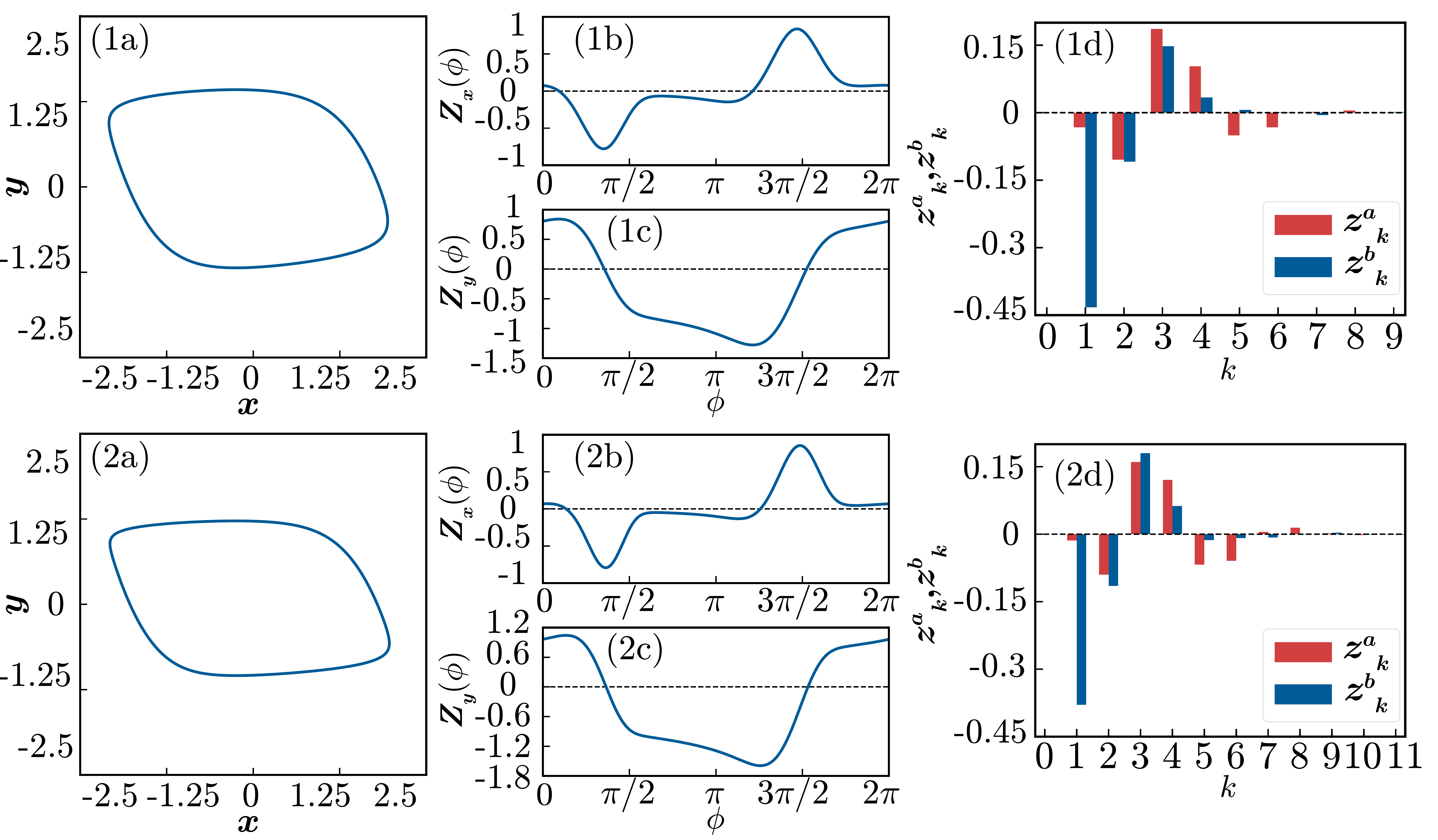}
 		\caption{
 			Limit cycle and phase sensitivity function of 
 			the FitzHugh-Nagumo model.
 			(1a, 2a): Limit cycle ${\bm X}_0(\phi)$.
 			\;
 			(1b, 2b): $x$ component $Z_x(\phi)$ of the PSF ${\bm Z}(\phi)$.
 			\;
 			(1c, 2c): $y$ component $Z_y(\phi)$ of the PSF ${\bm Z}(\phi)$.
 			\;
 			(1d, 2d): Fourier coefficients $z^a_k$ and $z^b_k$ of $Z_x(\phi)$.
 			\;
 			Figures (1a), (1b), (1c), and (1d) are for the parameter set (A), and (2a), (2b), 2(c), and 2(d) are for (B), respectively.}
 		\label{fig_3}
 	\end{center}
 \end{figure}

As an example of the limit cycle, we consider the FitzHugh-Nagumo model \cite{fitzhugh1961impulses, nagumo1962active},
\begin{align}
\label{eq:fhz}
\dot{x} &= x - a_1x^3 - y + u(\omega_e t),
\cr
\dot{y} &= \eta_1 (x + b_1),
\end{align}
where $\bm{X} = (x,y)^\mathsf{T}$ is the system state 
consisting of a voltage-like (activator) variable $x$
and a recovery-like (inhibitor) variable $y$, respectively,
$u(\omega_e t)$ is a periodic input with frequency $\omega_e$, 
and $(a_1, b_1, \eta_1)$ are parameters. 
We use two sets of parameters,
(A)  $(a_1, b_1, \eta_1) = (1/3, 0.25, 0.25)$ and 
(B)  $(a_1, b_1, \eta_1) = (1/3, 0.25, 0.15)$,
where the values of the parameter $\eta_1$ determining the timescale of $y$ are different while the values of the other parameters $a_1$ and $b_1$ are the same.
The limit cycle, PSF, and Fourier coefficients of the PSF 
for these parameter set are shown in Fig.~\ref{fig_3}.
The natural frequencies are $ \omega  =  0.404$ for (A) and $\omega = 0.286$ for (B), respectively.
We assume that the phase increases when the oscillator rotates in the counterclockwise direction on the $xy$ plane.
With $\delta = 0.001$ used here, the numbers of Fourier coefficients are $k_{max} = 9$ and $11$ for (A) and (B), respectively.
\begin{figure} [!t]
\begin{center}
\includegraphics[width=1.0\hsize,clip]{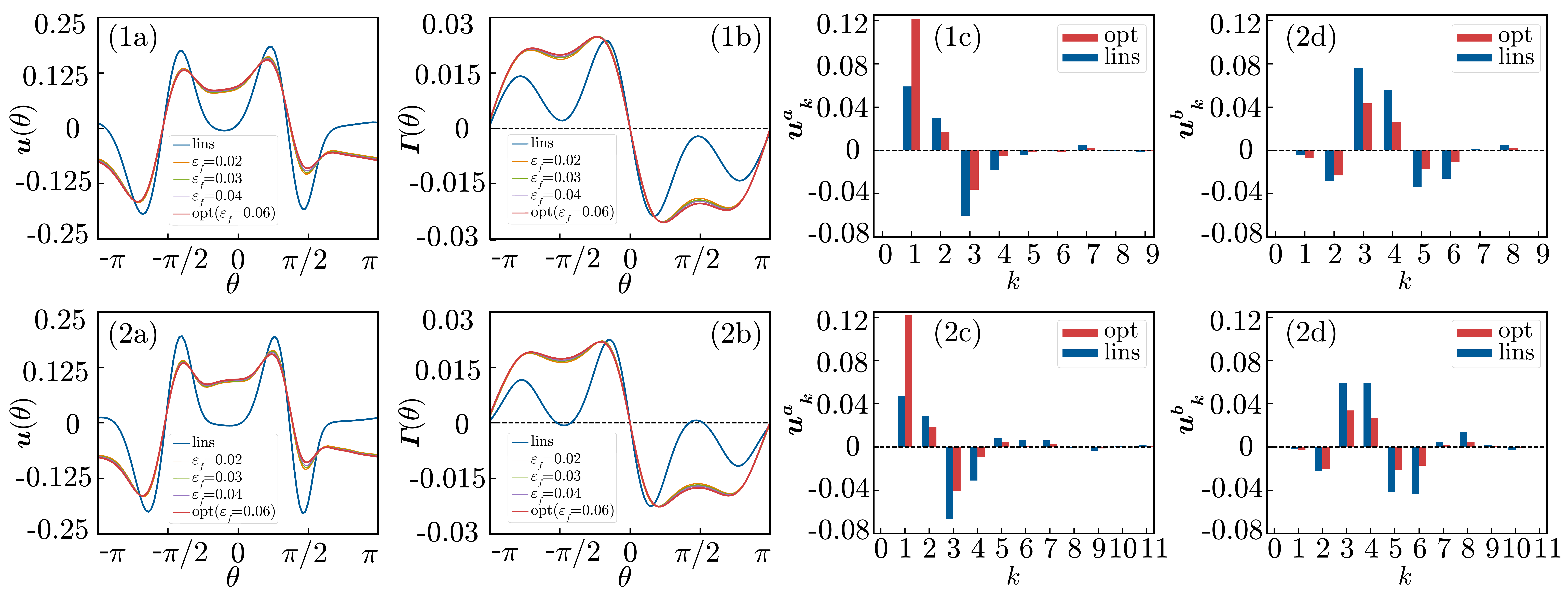}
\caption{
Optimal waveforms and phase coupling functions for achieving fast 
global convergence ($\epsilon_f = 0.06$) ("opt", red) and for improving local linear stability ("lins", blue).
(1a, 2a): Optimal waveforms $u$ of the periodic inputs.
(1b, 2b): Phase coupling functions $\Gamma$.
(1c, 2c): Fourier coefficients $u^a_k$ of the optimal waveforms.
(1d, 2d): Fourier coefficients $u^b_k$ of the optimal waveforms.
In (1a, 2a) and (1b, 2b), the results with $\epsilon_f = 0.02$ (orange), $0.03$ (light-green), and $0.04$ (purple) are also shown.}
\label{fig_4}
\end{center}
\end{figure}

In the numerical simulations, we use both parameter sets (A) and (B) and 
fix the input power as $P = 0.01$. We also assume $\Delta_e = 0$, that is, the frequency of the periodic input is identical with the frequency of the system, i.e., $\omega_{e} = \omega$, for simplicity. We numerically confirmed that qualitatively similar results to those presented in the next subsection are obtained also with small non-zero $\Delta_e$.

\subsection{Results}

In the numerical optimization of the objective function, we set the small parameters at $\epsilon_f = 0.06$ and $\epsilon_c = 0.001$,
respectively, and used the average of the convergence times from 100 uniformly distributed initial points $\psi_0$ in $\mathcal{A}_\Gamma$
to approximate $T_{ave}$.
We use the penalty method to take into account the constraints in Eq.~(\ref{eq:opt_obj_gcs}), namely, we add a large penalty to the objective function when these constraints are violated.
As the problem is non-convex, we use multiple initial guesses to find the optimal waveforms using the numerical solver.
As we set the Fourier coefficients randomly as the initial guess, the corresponding phase coupling function $\Gamma$ can have more than two pairs of stable and unstable fixed points as opposed to the assumption. To avoid such cases and ensure that only a single pair of stable and unstable fixed points exist, we also added a large penalty when $\Gamma$ crosses the horizontal axis three times or more.

We first consider the case with the parameter set (A).
In this case, the optimized waveform obtained in the previous study~\cite{zlotnik2013optimal}, which maximizes the local linear stability of the entrained state, yields slower global convergence than a simple sinusoidal waveform.
Thus, whether the present formulation can yield a better result than the previous local optimization is of our interest. 

Figures~\ref{fig_4}(1a) and \ref{fig_4}(1b) show the optimal waveform $u$ and phase coupling function $\Gamma$ for achieving fast global convergence, and Figs.~\ref{fig_4}(1c) and \ref{fig_4}(1d) show the Fourier coefficients 
$u^a_k$ and $u^b_k$ of the optimal waveforms.
For comparison, the corresponding quantities for maximizing the local linear stability of the entrained state, 
namely, $u(\theta) = {Z_x}'(\theta)$ and $\Gamma(\theta) = \la Z_x(\theta + \phi) {Z_x}'(\phi) \ra_\phi$
obtained in Ref.~\cite{zlotnik2013optimal}, are also shown.
As we see in Fig.~\ref{fig_4}(1c) and \ref{fig_4}(1d), the optimal waveform for the global convergence has smaller high-harmonic Fourier coefficients than the optimal waveform for the local linear stability, which leads to faster global convergence to the entrained state.
This can be observed from the phase coupling functions in Fig.~\ref{fig_4}(1b) representing the velocity of the phase difference $\psi$; the red curve (global convergence) is mostly kept away from the horizontal axis and the variation of the phase difference $\psi$ does not become very slow except near the fixed points, while the blue curve (linear stability) comes very close to the horizontal axis not only near the fixed points but also in other locations and the variation of $\psi$ becomes very slow around such points, yielding larger convergence time.

In Figs.~\ref{fig_4}(1a) and \ref{fig_4}(1b) we also plot the optimal waveforms $u$ and phase coupling functions $\Gamma$ for $\epsilon_f = 0.02, 0.03$, and $0.04$ in addition to those for $\epsilon_f = 0.06$.
We can confirm that the optimal waveform depends only slightly on the value of $\epsilon_f$ in this range.
Varying the value of $\epsilon_c$ around $0.001$ also does not affect the resulting waveform (not shown).
It should also be noted that, in the limit $\epsilon_f \to 0$, the optimal waveform approaches the one for maximizing the linear stability, because the linear dynamics near the fixed point dominates the convergence time in this limit.

\begin{figure} [!t]
\begin{center}
\includegraphics[width=0.8\hsize,clip]{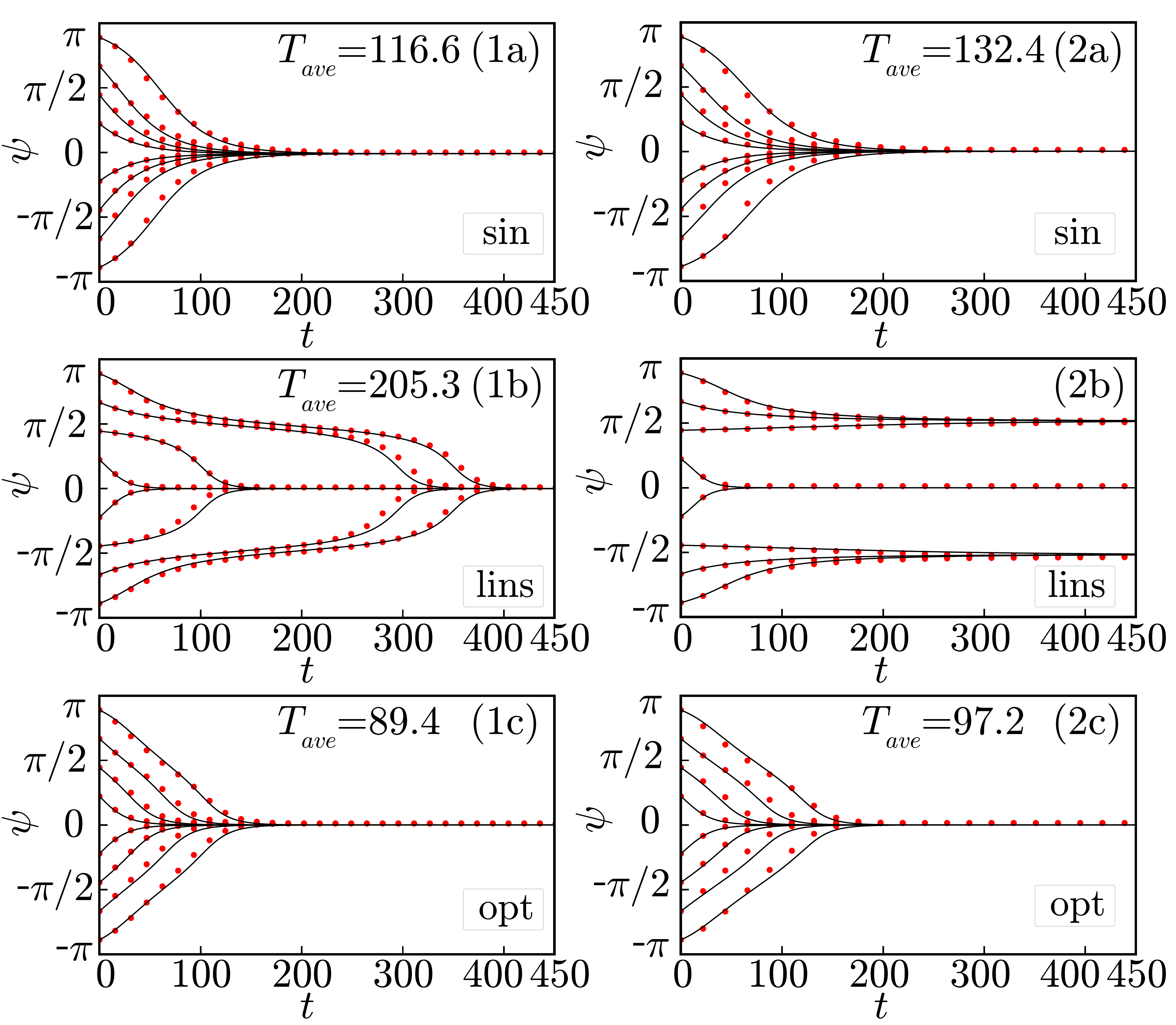}
\caption{
Convergence to the entrained state from $8$ initial values of 
the phase difference $\psi$.
The black lines represent the convergence of the phase difference $\psi$ 
described by Eq.~(\ref{eq:dpsi}) and 
the red dots represent the phase difference $\psi$ at $t=kT~(k=0, 1,\ldots)$ 
obtained from the direct numerical simulation of Eq.~(\ref{eq:X}) with Eq.~(\ref{eq:fhz}).
(1a, 2a): Sinusoidal waveform.
(1b, 2b): Optimal waveform for the local linear stability. 
(1c, 2c): Optimal waveform for the fast global convergence.
}
\label{fig_5}
\end{center}
\end{figure}

Figures~\ref{fig_5}(1a), \ref{fig_5}(1b), and \ref{fig_5}(1c) compare
the convergence of the phase difference $\psi$ to the entrained state 
for three types of input waveforms,
i.e., the sinusoidal waveform (1a), the optimal waveform for the linear stability (1b), and the optimal waveform for the fast global convergence (1c).
The black-thin lines represent the convergence of the phase difference $\psi$ from uniformly distributed $8$ initial points
obtained from Eq.~(\ref{eq:dpsi}) for the phase difference and the red dots represent the phase difference obtained by direct numerical simulations of Eq.~(\ref{eq:X}) with Eq.~(\ref{eq:fhz}), showing good agreement.
The average convergence times to the entrained state 
obtained by direct numerical simulations of Eq.~(\ref{eq:dpsi})
are $T_{ave} = 116.6$, $205.3$ and $89.4$
in Figs.~\ref{fig_5}(1a), \ref{fig_5}(1b), and \ref{fig_5}(1c), respectively.
Here, the averaged convergence time is evaluated by 
numerically integrating Eq.~(\ref{eq:dpsi}) from $100$ uniformly distributed initial points. 

The average convergence time is the smallest when the optimal waveform for the global convergence is used, while it takes the largest value when the optimal waveform for the local linear stability is used; the sinusoidal waveform gives an intermediate result.
Note that in Fig.~\ref{fig_5}(1b) with the optimal waveform for local linear stability,
the convergence from the initial points near $0$ is faster than the other two cases, but the convergence from other initial points far from $0$ is slower, resulting in a larger convergence time on average.

We next consider the case with parameter set (B). In this case, the optimal waveform for the local linear stability gives rise to several stable fixed points and does not achieve global convergence.

Figures~\ref{fig_4}(2a) and \ref{fig_4}(2b) show the optimal waveforms $u$ and the phase coupling functions $\Gamma$, and Figs.~\ref{fig_4}(2c) and \ref{fig_4}(2d) 
show the Fourier coefficients $u^a_k$ and $u^b_k$ of the periodic input waveforms, respectively, again for the optimization of global convergence and for the optimization of local linear stability.
Since we vary only the parameter $\eta_1$, the results for (B) are similar to those for (A), but the phase coupling function optimized for the local linear stability has three stable fixed points as seen in Fig.~\ref{fig_4}(2b).
In Figs.~\ref{fig_4}(2a) and \ref{fig_4}(2b), we also plot the optimal waveforms $u$ and phase coupling functions $\Gamma$ for $\epsilon_f = 0.02, 0.03$, and $0.04$ in addition to those for $\epsilon_f = 0.06$. We can confirm that the optimal waveform depends only slightly on the value of $\epsilon_f$ in this range.

Figures~\ref{fig_5}(2a), \ref{fig_5}(2b), and \ref{fig_5}(2c) show the convergence of $\psi$ to the entrained state for the cases with the sinusoidal waveform (2a), 
the optimal waveform for the linear stability (2b), 
and the optimal waveform for the global convergence (2c), respectively.
The average convergence time is smaller for the case with the optimal waveform for the global convergence than the case with the sinusoidal waveform.
The average convergence times from uniformly distributed $100$ initial points
calculated by direct numerical simulations of Eq.~(\ref{eq:dpsi})
are $T_{ave} = 132.4$ and $97.2$ in Figs.~\ref{fig_5}(2a) and (2c), respectively.
The global convergence is not achieved 
in Fig.~\ref{fig_5}(2b) in the case with the optimal waveform for the local linear stability.

Thus, we have confirmed that the present formulation of the optimal waveforms that takes into account the average convergence time can achieve faster global convergence to the entrained state than the previous optimal waveforms obtained by considering the local linear stability of the entrained state for the examples considered here. 

\section{Realizing prescribed phase distributions}
\label{optimization2}

As the second problem, we consider a population of identical uncoupled oscillators 
subjected to small noise and 
realize 
prescribed global phase distributions
by optimizing the input waveform.
This type of problem is important for controlling populations of oscillatory cells
in medical engineering applications, for example, 
cardiac pacemaker cells \cite{michaels1987mechanisms} 
and deep brain stimulation for the treatment
of Parkinson's disease \cite{tass2001desynchronizing}
and epileptic seizure \cite{good2009control}.
In preceding studies, Lyapunov-based control \cite{monga2018synchronizing, kuritz2019ensemble}
and optimal control \cite{monga2019phase2, wilson2020optimal} of oscillator populations have been analyzed.
Control of the oscillator population by using common noisy forcing has also been analyzed~\cite{kurebayashi2014design}.
Our method presented here gives a simpler formulation of the optimization of the input waveform by using the averaging procedure \cite{kuramoto1984chemical} and explicitly incorporates the effect of small noise.

\subsection{Formulation of the optimization problem}

We consider a large population of identical uncoupled oscillators subjected to a common periodic input and independent Gaussian-white noise, both of which are weak.
By regarding the input and noise as perturbations, we can apply the phase reduction
and formulate the optimization problem as in Eq.~(\ref{eq:opt_obj2}) also in this case.
We assume that each oscillator in the population independently obeys an Ito stochastic differential equation (SDE),
\begin{align}
\label{eq:X2}
d{\bm{X}}(t) = \{ {\bm{F}}({\bm X}(t)) +  
u(\omega_e t)(1, 0, \ldots, 0)^\mathsf{T} \} dt 
+ {\bm{G}}({\bm X}(t)) d {\bm W}(t),
\end{align}
where $u(\omega_e t)$ is the periodic input force and $\bm{W}(t) = (W_1(t), \ldots, W_N(t))^\mathsf{T} \in \mathbb{R}^{N \times 1}$ 
is a vector of independent Wiener processes $W_{i}(t) (i=1, \ldots, N)$ satisfying 
$\mathbb{E}[{dW_{i}(t)dW_{j}(t)}] = \delta_{ij} dt$ 
and ${\bm{G}}({\bm X}) \in {\mathbb R}^{N \times N}$ 
represents the noise intensity matrix.
The periodic input $u$ is assumed weak and of $O(\varepsilon)$ as before, and each component of ${\bm G}({\bm X})$ is also assumed weak and of $O(\sqrt{\varepsilon})$.

From Eq.~(\ref{eq:X2}), by using the phase reduction method for limit-cycle oscillators driven by white noise~\cite{kato2019semiclassical, kato2020semiclassical, aminzare2019phase}, we can derive a SDE for the phase $\phi = \Phi({\bf X})$ of the oscillator as
\begin{align}
\label{eq:dphi2}
d\phi &= \left\{ \omega +  \bm{Z}( \phi ) \cdot u(\omega_e t)(1, \ldots, 0)^\mathsf{T} + g(\phi) \right\} dt + \{ \bm{G}(\phi)^\mathsf{T} {\bm Z}(\phi) \} \cdot d\bm{W}.
\end{align}
Here, $g(\phi) = \frac{1}{2} \mbox{Tr} \left\{ {\bm G}(\phi)^\mathsf{T} {\bm Y}(\phi) {\bm G}(\phi) \right\}$ is a term arising from the Ito formula, where  ${\bm Y}(\phi) = \nabla^\mathsf{T} \nabla \Phi|_{{\bm X} = {{\bm X}_0(\phi) }} \in {\mathbb R}^{N \times N}$ is a 
Hessian matrix of the phase function $\Phi({\bm X})$ 
evaluated at ${\bm X} = {\bm X}_0(\phi)$ on the limit cycle.
By further applying the averaging procedure 
as explained in Ref.~\cite{kato2019semiclassical}, we can derive an approximate equation for the phase difference $\psi = \phi - \omega_e t$ as
\begin{align}
\label{eq:dpsi2}
d\psi = \left\{ \tilde{\Delta}_e + \Gamma(\psi) \right\} dt
+ \sqrt{2 D_0} d W,
\end{align}
where
$ \tilde{\Delta}_e = \tilde{\omega} - \omega_e$
is the detuning of the effective oscillator frequency 
$\tilde{\omega} := \omega + \la g(\theta) \ra_{\theta}$
from the frequency $\omega_e$  of the periodic modulation,
$\sqrt{2 D_0} = \sqrt{\la \sum_{i} \left( \bm{G}(\theta)^\mathsf{T} {\bm Z}(\theta) \right)^2_i \ra_\theta}$ is the effective intensity of the noise, and $W(t)$ is a Wiener process satisfying $\mathbb{E}[{dW(t)dW(t)}] = dt$.
The PSF ${\bm Z}$ \cite{brown2004phase} and the Hessian matrix ${\bm Y}$\cite{suvak2010quadratic} can be numerically obtained as $2 \pi$-periodic solutions to linear adjoint-type equations with appropriate constraints  (see also the Appendix B in Ref.~\cite{kato2019semiclassical}). 

The Fokker-Planck equation (FPE) for the probability density function (PDF) $P(\psi, t)$ of the phase difference $\psi$ described by the SDE~(\ref{eq:dpsi2}) is given by
\begin{align}
\label{eq:fpephi}
\frac{\partial} {\partial t} P(\psi, t) =
- \frac{\partial}{\partial \psi} 
\left[
 \left\{ \tilde{\Delta}_e + \Gamma(\psi) \right\} P(\psi, t) \right] 
+ D_0\frac{\partial^2}{\partial \psi^2}  P(\psi, t),
\end{align}
which has a steady-state PDF given by
\begin{align}
\label{eq:std_prob}
P_s(\psi) &=\frac{1}{C} \int_{\psi}^{\psi+2 \pi} \exp \left(\frac{v\left(\psi^{\prime}\right)-v(\psi)}{D_0}\right)
d\psi^{\prime},
& v(\psi) = - \int^{\psi}_0 \{ \tilde{\Delta}_e  + \Gamma(\theta) \} d\theta,
\end{align}
where $C$ is a normalization constant \cite{pikovsky2001synchronization}.
Note that $v(\psi)$ defined above can be regarded as a potential for the deterministic part of Eq.~(\ref{eq:dpsi2}).
Our aim is to optimize the input waveform so that the steady-state PDF becomes closest to the target PDF.

To this end, we 
write 
the steady-state PDF 
$P_s(\psi)$ and target PDF $\tilde{P}_s(\psi)$
in Fourier series as
\begin{align}
P_s(\psi) &= \frac{1}{2\pi} + \sum_{k = 1}^{\infty} ( p^a_{k} \cos k \psi + p^b_{k} \sin k\psi ),~ 
\cr
\tilde{P}_s(\psi) &= \frac{1}{2\pi} + \sum_{k = 1}^{\infty} ( \tilde{p}^a_{k} \cos k \psi + \tilde{p}^b_{k} \sin k \psi ), 
\end{align}
where
\begin{align}
p^a_k(\psi) &= \frac{1}{\pi} \int_{0}^{2\pi}  P_s(\psi) \cos k \psi d\psi,~ 
p^b_k(\psi) = \frac{1}{\pi} \int_{0}^{2\pi}  P_s(\psi) \sin k \psi d\psi, 
\cr
\tilde{p}^a_k(\psi) &= \frac{1}{\pi} \int_{0}^{2\pi}  \tilde{P}_s(\psi) 
\cos k \psi d\psi,~ 
\tilde{p}^b_k(\psi) = \frac{1}{\pi} \int_{0}^{2\pi}  
\tilde{P}_s(\psi) \sin k \psi d\psi,
\end{align}
and introduce a distance $D_{prob}$ between the two PDFs as
\begin{align}
\label{eq:dist_prob}
D_{prob} = 
\sqrt{ \sum_{k = 1}^{\infty} \left[ ( \tilde{p}^a_k -  p^a_k)^2 + ( \tilde{p}^b_k -  p^b_k)^2 \right]}.
\end{align}
The optimization problem (\ref{eq:opt_obj2}) is then formulated as follows:
\begin{align}
\label{eq:opt_obj_prob}
&\underset{u^a_0, u^b_0, u^a_1, u^b_1, \cdots}{\mbox{minimize}}\ 
\sqrt{ \sum_{k = 1}^{\infty} \left[ ( \tilde{p}^a_k -  p^a_k)^2 + ( \tilde{p}^b_k -  p^b_k)^2 \right]}
&\quad \mbox{s.t.} \quad
\frac{(u^{a}_0)^2}{4} + \frac{1}{2} \sum_{k = 1}^{\infty} \left( (u^{a}_k)^2 +  (u^{b}_k)^2 \right) = P.
\end{align}

Here, we note the difference from the deterministic problem considered in the previous section where we had to exclude the neighborhoods of the fixed points in order to avoid divergence of the convergence time.
In the present case, because each oscillator is subjected to Gaussian-white noise, the phase difference $\psi$ can reach any value on $[-\pi, \pi]$ from arbitrary initial values in the long run. Also, we do not consider the time necessary for the convergence but rather focus only on the final steady-state distribution of the phase difference, where the convergence to the final steady-state distribution from any initial distribution is guaranteed by the H-theorem for the Fokker-Planck equation~\cite{horsthemke2006noise}.
Thus, we do not need to restrict the initial and final states of the oscillators in the present stochastic case.

\subsection{Noisy FitzHugh-Nagumo Model}

As an example of the limit cycle, we use the FitzHugh-Nagumo model~\cite{fitzhugh1961impulses, nagumo1962active} as before,
but whose $x$ variable is now subjected to small noise, described by the following SDE:
\begin{align}
\label{eq:fhz2}
dx &= \{ x - a_1x^3 - y + u(\omega_e t) \} dt + \sqrt{2 D_1} dW,
\cr
dy &= \eta_1 (x + b_1) dt,
\end{align}
where $dW$ is the Wiener process and $D_1$ represents the intensity of the weak noise 
( $0 \leq D_1 \ll 1$). 
We only consider the case with the parameter set (B) in Sec.~\ref{opt1_model}
and set the noise intensity as $D_1 = 0.01$. The effective frequency of the oscillator
under the effect of the noise is evaluated as $\tilde{\omega}  = 0.288$ in this case ($\omega  = 0.286$ in the absence of noise).
As in Sec.~III, 
we also assume $\tilde{\Delta}_e = 0$, that is, the frequency of the periodic input is identical with the frequency of the system, $\tilde{\omega}_{e} = \omega$. We numerically confirmed that qualitatively similar results to those 
in the following subsections are obtained also for small non-zero $\tilde{\Delta}_e$.

\subsection{Results}

The cluster state of the oscillator population refers to the state in which the oscillators form several groups with different phase values. In Refs.~\cite{matchen2017real, tass2003model}, it is argued that the formation of cluster states can be effective in desynchronizing pathologically synchronized neurons in Parkinson's disease by deep brain stimulation.
Here, as the target phase distributions, we consider (1) two-cluster and (2) three-cluster states of the oscillators given by the following weighted sums of the von Mises distributions~\cite{mardia2009directional}:
\begin{align}
\label{eq:des_prob}
\tilde{P}^1_s(\psi) &= 
\frac{
e^{\kappa \cos \left( \psi - \pi/2 \right)}
+
3e^{\kappa \cos \left( \psi + \pi/2 \right)}
}
{8 \pi \mathcal{I}_{0}(\kappa)},
\cr
\tilde{P}^2_s(\psi) &= 
\frac{
3 e^{\kappa \cos \psi }
+
e^{\kappa \cos \left( \psi - 2\pi/3 \right)}
+
e^{\kappa \cos \left( \psi + 2\pi/3 \right)}
}
{10 \pi \mathcal{I}_{0}(\kappa)},
\end{align}
where $\mathcal{I}_{0}(\kappa)$ is the modified Bessel function of 
the first kind of order $0$ and we set $\kappa = 5$ in the following calculations.
Since the problem is non-convex, we use multiple initial guesses to obtain the optimal Fourier coefficients by the numerical solver also in this case.
The maximum number of the Fourier coefficients used in the numerical calculation is $k_{max} = 11$.
We set the input power as $P=0.002$ and $0.001$ for the cases (1) and (2), respectively.

\begin{figure} [!t]
\begin{center}
\includegraphics[width=1.0\hsize,clip]{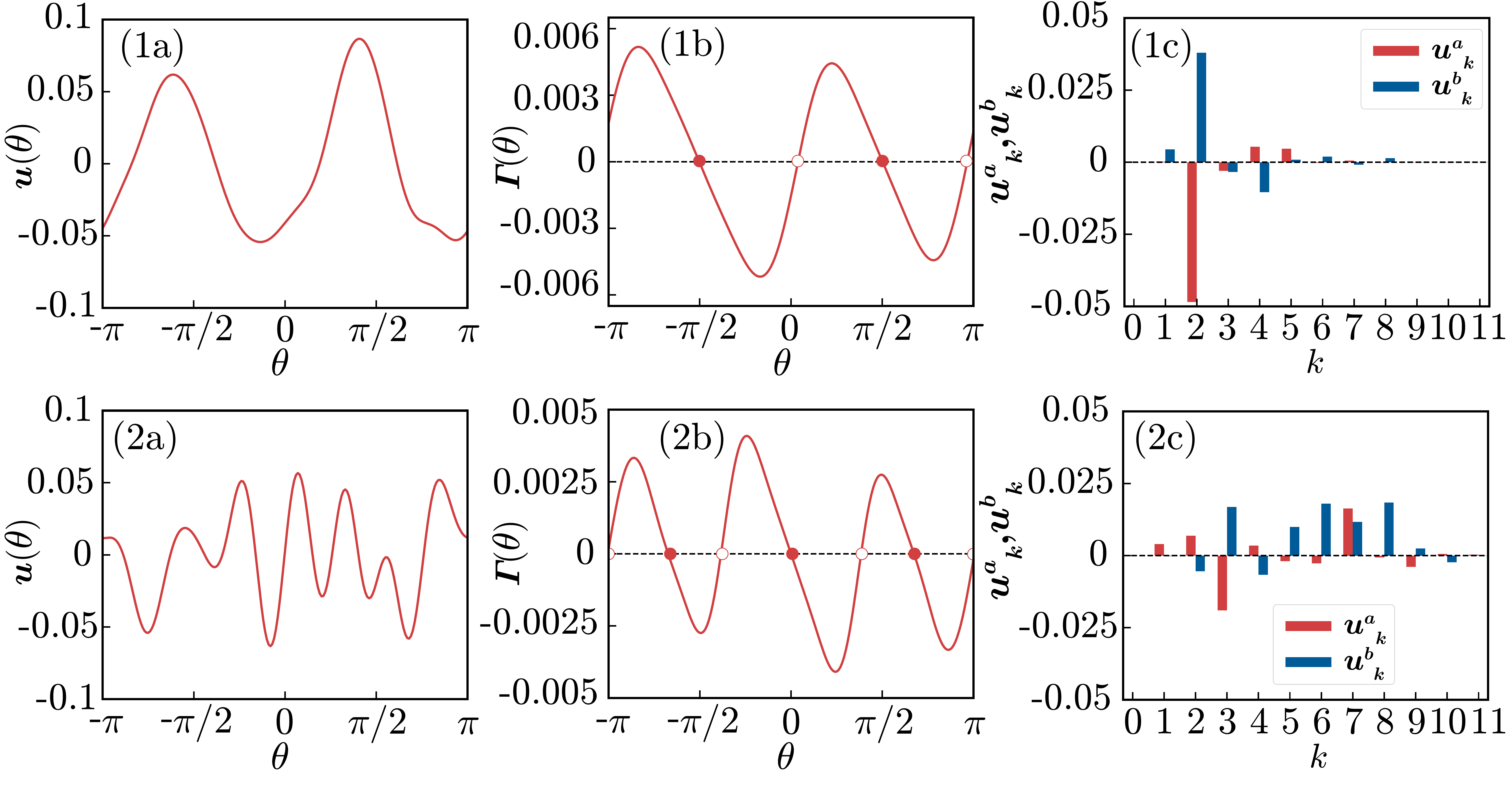}
\caption{
Optimal waveforms and phase coupling functions for realizing 
(1) two-cluster and (2) three-cluster states.
(1a, 2a): Optimal waveforms $u$ of the periodic inputs.
(1b, 2b): Phase coupling functions $\Gamma$.
(1c, 2c): Fourier coefficients $u^a_k$ (red) and $u^b_k$ (blue) of the optimal waveforms.
}
\label{fig_6}
\end{center}
\end{figure}

\begin{figure} [!t]
\begin{center}
\includegraphics[width=1.0\hsize,clip]{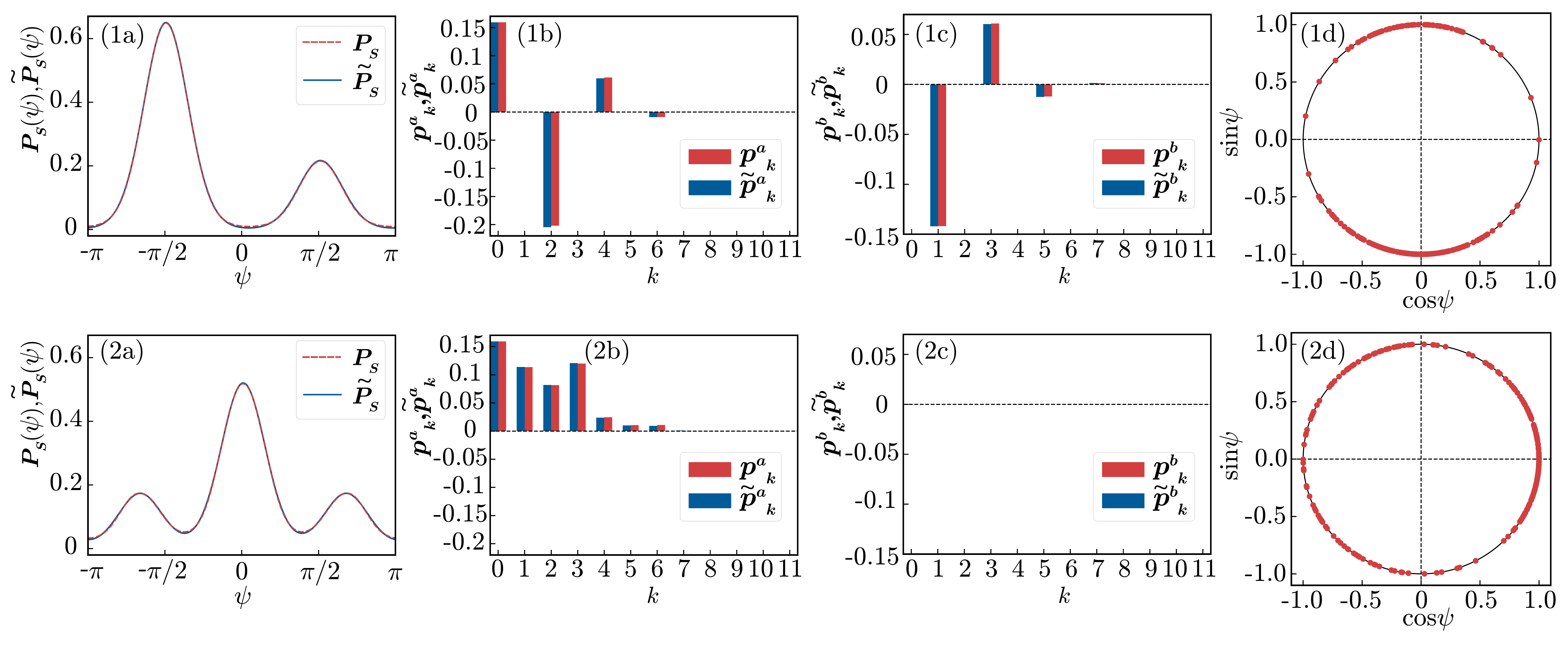}
\caption{
Realization of weighted cluster states.
(1a,2a): 
Steady-state PDFs $P_s(\psi)$ (red) and
target PDFs $\tilde{P}_s(\psi)$ (blue).
(1b,2b): Fourier coefficients $p^a_k$ (red) and $\tilde{p}^a_k$ (blue) of $P_s(\psi)$ and $\tilde{P}_s(\psi)$.
(1c,2c): Fourier coefficients $p^b_k$ (red) and $\tilde{p}^b_k$ (blue) of $P_s(\psi)$ and $\tilde{P}_s(\psi)$.
(1d,2d): Snapshots of the phase differences of 250 oscillators at time $t=500T$
obtained by direct numerical simulations of Eq.~(\ref{eq:fhz2}).
}
\label{fig_7}
\end{center}
\end{figure}

Figures~\ref{fig_6}(1a)-\ref{fig_6}(1c) and \ref{fig_6}(2a)-\ref{fig_6}(2c)
show the optimal waveforms $u$,
phase coupling functions $\Gamma$, 
and the Fourier coefficients $u^a_k$ and $u^b_k$ of the optimal waveforms
for the target PDFs (1) and (2), respectively.
As we see in Figs.~\ref{fig_6}(1b) and (2b), the realized $\Gamma$ has stable fixed points near the peaks of the target distributions (see Figs.~\ref{fig_7}(1a) and (2a)).
We can also observe that the fixed point corresponding to a higher peak of the target PDF 
takes a lower value of the potential $v(\psi)$ defined in Eq.~(\ref{eq:std_prob}), as can be seen from 
the larger difference in height between the peaks of $\Gamma$ at both sides of the fixed point.
In Figs.~\ref{fig_6}(1c) and (2c), the Fourier coefficients for $k=2$ and $3$ take large values for the (1) two-cluster and (2) three-cluster cases, respectively, as expected.  In the three-cluster case, we also observe non-small higher-harmonic Fourier coefficients at larger values of $k$, which are used to reproduce the target distributions precisely. These higher-harmonic components can be eliminated by decreasing the input power, but it also degrades the reproducibility of the target PDF.


Figures~\ref{fig_7}(1a)-\ref{fig_7}(1c) and \ref{fig_7}(2a)-\ref{fig_7}(2c)
show the optimized steady-state PDFs $P_s(\psi)$, the target PDFs $\tilde{P}_s(\psi)$, and their Fourier coefficients $p^a_k$, $\tilde{p}^a_k$, $p^b_k$ and $\tilde{p}^b_k$
for the target PDFs (1) and (2), respectively.
The realized steady-state PDFs $P_s(\psi)$
are very close to the target PDFs $\tilde{P}_s(\psi)$
and virtually indistinguishable from the target PDFs,
where the distance between the realized and target PDFs
are $D_{prob} = 0.00328$ and $0.00310$
for the target PDFs (1) and (2), respectively.
Note that the Fourier coefficients $p^b_k$ and $\tilde{p}^b_k$ in Fig.~\ref{fig_7}(2c) 
vanish because
the target PDF $\tilde{P}^2_s(\psi)$ is an even function. 
In Figs.~\ref{fig_7}(1d) and (2d), we plot the phase difference $\psi$ of $250$ oscillators at time $t = 500T$ obtained by direct numerical simulations of
the stochastic dynamics given by Eq.~(\ref{eq:X2}) with Eq.~(\ref{eq:fhz2}) using the optimal waveforms
for the target PDFs (1) and (2), respectively, 
where the initial oscillator states are uniformly distributed on the limit cycle. 

We can observe that the oscillator population indeed exhibits 
a two-cluster or three-cluster state, 
where the numbers of oscillators in the clusters are $N = 72~(0 \leq \psi < \pi)$ and $178~(-\pi \leq \psi < 0)$ for the target PDF (1), and $N = 149~(-\pi/3 \leq \psi < \pi/3), 56~(\pi/3 \leq \psi < \pi)$, and $45~( -\pi \leq \psi < -\pi/3)$ for the target PDF (2), respectively. 
Thus, we have confirmed that the present formulation of the optimal waveforms can successfully realize the target PDFs for the examples considered here.

In Fig.~\ref{fig_7}, we could precisely reproduce the target PDFs and attain small values of $D_{prob}$ because the target PDFs consisting of the sums of von Mises distributions are sufficiently smooth and close to the PDFs that can be represented as stationary solutions of the FPE (\ref{eq:std_prob}).
For a more general class of target PDFs with higher-harmonic Fourier components, 
e.g., strongly sharp-edged von Mises distributions with large $\kappa$, the optimal waveform may not reproduce the target PDFs very well.
From Eq.~(\ref{eq:std_prob}), the sharpness of the reproduced PDF is limited by the steepness of the potential $v(\psi')$ and noise intensity $D_0$, determined mainly by $P, D_0$, and $k_{max}$.
%
In the next subsection, we discuss the reproducibility of PDFs with high-harmonic Fourier components.
\begin{figure} [!t]
\begin{center}
\includegraphics[width=0.85\hsize,clip]{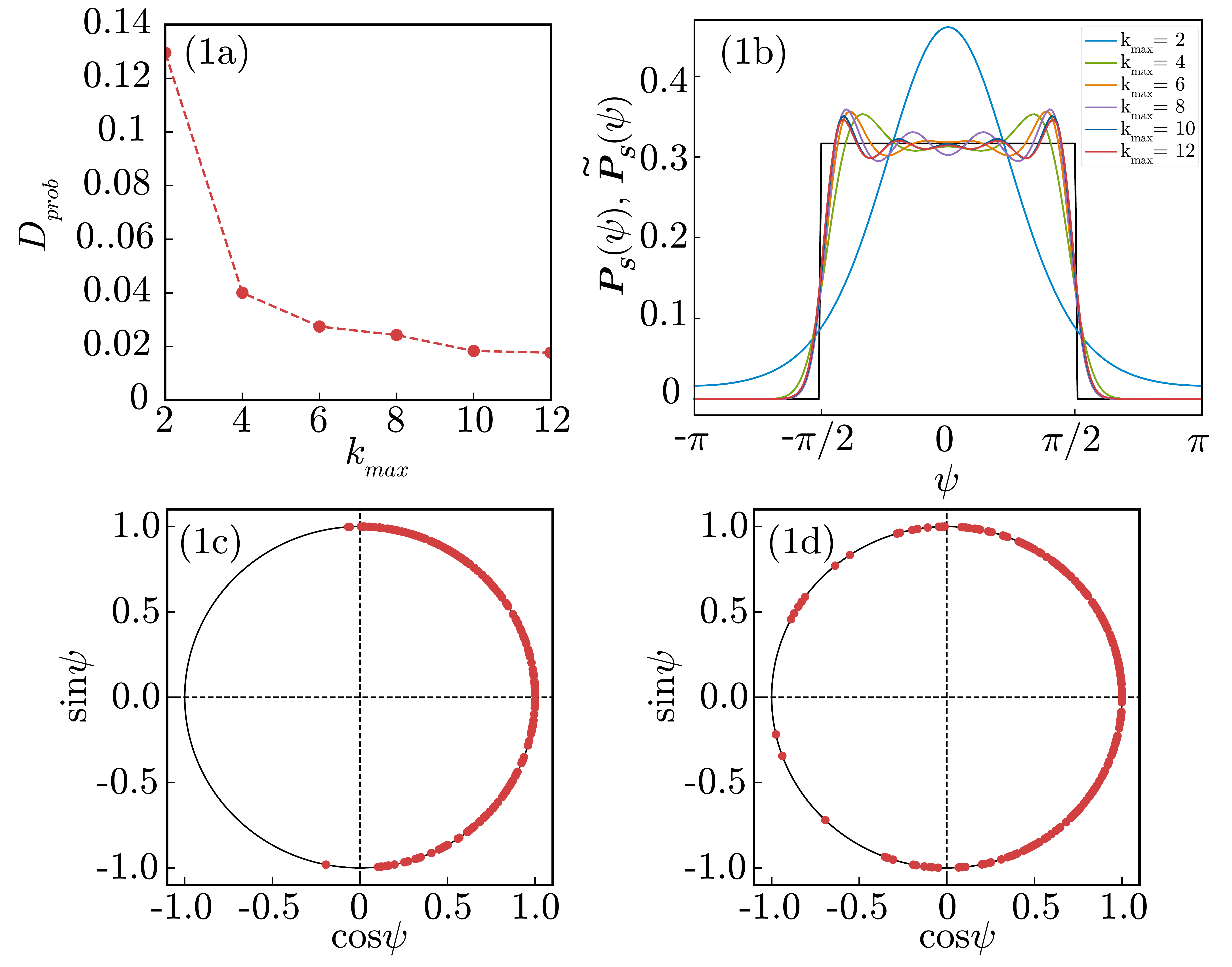}
\caption{
Reproducibility of a uniform PDF on a half interval of the oscillator phase.
(1a): Distance $D_{prob}$ when varying the number of the Fourier coefficients $k_{max}$ to be optimized.
(1b): Steady-state PDFs $P_s(\psi)$ (colored lines) and
the target PDF $\tilde{P}_s(\psi)$ (black line).
(1c, 1d): Phase differences of 250 oscillators at time $t = 400T$ obtained by direct numerical simulations of Eq.~(\ref{eq:X2}) and Eq.~(\ref{eq:fhz2}) when (1c) $k_{max} = 12$ and (1d) $k_{max} = 2$.
}
\label{fig_8}
\end{center}
\end{figure}

\subsection{Reproducibility of a uniform distribution}

To analyze the reproducibility of PDFs with high-harmonic Fourier components, we here consider the following uniform PDF on a half interval of the phase difference, 
\begin{align}
\tilde{P}^3_s(\psi) &=  
\left\{\begin{array}{cl}
\frac{1}{\pi} \quad & \left(- \frac{-\pi}{2}  \leq \psi  \leq \frac{\pi}{2}\right), \\
0 \quad  & \left( -\pi  \leq \psi  <  \frac{-\pi}{2},  \frac{\pi}{2}  <  \psi  <\pi  \right), \\
\end{array}\right.
\end{align}
whose Fourier coefficients are given by $p^a_k = \frac{2}{k \pi^2} \sin \frac{k \pi}{2}$ and $p^b_k = 0~(k=1,2,\ldots)$. Note that the decay of $p_k^a$ with $k$ is slow (algebraic).
We set the input power as $P=0.01$.

Figure \ref{fig_8} compares the target PDF $\tilde{P}^3_s(\psi)$ and the realized PDFs ${P}_s(\psi)$ for differing values of the maximum number $k_{max}$ of the Fourier coefficients used in the numerical optimization.
As seen in Figs.~\ref{fig_8}(a) and (b), the distance $D_{prob}$ between the two PDFs 
decreases and the approximation of the target PDF $\tilde{P}^3_s(\psi)$ improves as $k_{max}$ is increased.
In Figs.~\ref{fig_8}(c) and (d), snapshots of the phase differences $\psi$ of $250$ oscillators at time $t = 400T$ obtained by direct numerical simulation of the dynamics in Eq.~(\ref{eq:fhz2}) 
 from uniformly distributed initial oscillator states on the limit cycle are shown for the cases with $k_{max} = 12$ and $k_{max} = 2$, respectively.
We can see that the 
oscillators successfully form a uniform distribution on a half interval of the phase when $k_{max} = 12$ within a small error, but they fail to do so when $k_{max} = 2$ and
several oscillators are considerably out of the range
$\left(- \frac{-\pi}{2}  \leq \psi  \leq \frac{\pi}{2}\right)$.

As we discussed before, the reproducibility of the target PDF has a fundamental limitation and may not be improved even if $k_{max}$ is further increased.
This is because the high-harmonic components of the PSF of the
FHN model in the present case
practically 
vanish above the wavenumber $k=10$,
and addition of harmonic components higher than this wavenumber does not contribute to the phase coupling function $\Gamma$.
The additive noise of effective intensity $D_0$ applied to the phase difference also introduces a minimum length scale, below which the stationary PDF is smoothed out and does not faithfully reproduce the structures of the target PDF.
Also, even if the PSF of the oscillator has very high-harmonic components, 
the numerical cost for solving the nonlinear programming can be considerably
large.
Thus, in practice, we need to choose an appropriate value of $\delta$ or $k_{max}$ by considering the balance between the precision of approximation and the computation time.

\subsection{Comparison with other methods for population control}

In this section, our objective is to maintain steadily
the target PDF in a noisy environment and thus we need to apply a periodic input continuously to the oscillator.
In the preceding studies, the Lyapunov-based control~\cite{monga2018synchronizing, kuritz2019ensemble} and optimal control~\cite{monga2019phase2} for realizing given target PDFs of the oscillator phase have been proposed.
When the noise is absent and the frequencies of the oscillators are identical, the PDF of the oscillator phase does not change its shape but simply rotates on $[0, 2\pi]$. Therefore, these control methods do not require the control power once the target PDF is realized and therefore they are more energy-efficient than the method proposed here in such situations. When the noise or frequency heterogeneity exists, continuous application of the control input is required also in these methods.

Another difference of the present method from the above preceding methods is that the present method gives a feedforward control and does not require the monitoring of the system state. It is therefore easier to implement than the preceding methods, which are of feedback type and require continuous monitoring of the system state.
It should also be noted that the present method is applicable even if the oscillators are mutually coupled or the oscillator frequencies are inhomogeneous, as long as their effects are sufficiently weak and the input waveform can dominate the oscillator dynamics.
Our method may not work well for a population of {\it coupled oscillators} undergoing collectively synchronized oscillations, where the coupling intensity is relatively strong and comparable to the control input.
For such systems, a feedforward open-loop control method for the coupled oscillator population is proposed in Ref.~\cite{wilson2020optimal}, which uses the phase-amplitude reduction framework to deal with their collective dynamics. 
Detailed comparison of the relative merits and power efficiency of these models would be important in future applications.


\section{Concluding remarks}
In this paper, we proposed a general optimization method of the periodic input waveform for global entrainment of weakly forced limit-cycle oscillators using phase reduction and nonlinear programming.
We showed that our method gives a simple feedforward control that can achieve fast global convergence of the oscillator to the entrained state and realize prescribed weighted cluster states of the oscillator population.
Our method can be formulated for a wide class of oscillator states and can be solved with commonly-used numerical tools for nonlinear programming. The objectives and constraints arising in the practical experimental setups can also be included in the optimization problems.
Thus, it may be practically applicable in various fields of science and technology in which control of rhythmic dynamics is required.

A main limitation in the present method is the assumption of the weakness of the forcing input.
Because the proposed method is based on the phase reduction theory, it is applicable only when the deviation of the oscillator state from the unperturbed limit cycle is small, namely, the periodic input should be sufficiently weak.
To cope with this problem, it will be helpful to include minimization of the deviation from the limit cycle in the objective function, which can be characterized by introducing the {\it isostable} or {\it amplitude} coordinates of the limit-cycle oscillator \cite{mauroy2013isostables, wilson2016isostable, shirasaka2017phase, shirasaka2020phase, wilson2018greater, monga2019optimal, wilson2021optimal, takata2021fast} in addition to the phase coordinate.
Such an extension will allow us to use stronger forcing input and thereby achieve even faster global entrainment.

In the present study, we considered a single-oscillator problem and optimized the input waveform.
In our previous study~\cite{shirasaka2017optimizing, watanabe2019optimization}, optimization of mutual coupling 
between a pair of coupled limit-cycle oscillators has also been analyzed by using the phase reduction and calculus of variations.
The present method can be readily extended to such optimization problems for the synchronization of coupled oscillators  and can be used to deal with a wide range of optimization problems that are unsolvable by using the calculus of variations.
Furthermore, it will also be interesting to apply the present method to quantum limit-cycle oscillators in the semiclassical regime. In our previous study~\cite{kato2019semiclassical, kato2020semiclassical},  semiclassical phase reduction theory for quantum limit-cycle oscillators has been developed and 
optimization problems of the waveform for maximizing linear stability of the entrained state and phase coherence of the oscillator	
have been analyzed by using the calculus of variation. 
By applying the present method of realizing weighted cluster states, we may also be able to realize weighted cluster states of quantum limit-cycle oscillators in the semiclassical regime.

\acknowledgments{We acknowledge JSPS KAKENHI JP17H03279, JP18H03287, JPJSBP120202201, JP20J13778, and JST CREST JP-MJCR1913 for financial support. }

\section*{Declarations}
\textit{Conflict of interest }
The authors declare that they have no conflict of interest.

\textit{Data availability}
All data generated or analysed during this study are included in this published article.



\appendix

\section{Reformulation of the optimization problems in previous works}
In this Appendix, we reformulate the optimization problems for the entrainment of a periodically forced limit-cycle oscillator, which have been treated by the calculus of variations in Refs.~\cite{harada2010optimal, zlotnik2013optimal, pikovsky2015maximizing}, by using the proposed method and obtain the optimal waveforms.

Here, we assume that the effective detuning is $\tilde{\Delta}_e = 0$ and that the entrained state is $\psi = 0$ without loss of generality.
The optimization problems for the improvement of maximum entrainment range (in a restricted sense) \cite{harada2010optimal,  zlotnik2012optimal, zlotnik2011optimal},
entrainment stability \cite{zlotnik2013optimal}, and enhancement of phase coherence \cite{pikovsky2015maximizing} 
are written as
\begin{align}
\label{opt_obj_ml}
\mbox{maximize}~ \int_{0}^{2\pi} \left( -Z_1(\theta) \right) u(\theta) d\theta, ~ \mbox{s.t.}~\la u^2 (\theta)\ra_{\theta} = P,
\end{align}
\begin{align}
\label{opt_obj_fe}
\mbox{maximize}~ \int_{0}^{2\pi} \left( -{Z}'_1(\theta) \right) u(\theta) d\theta, ~ \mbox{s.t.}~\la u^2 (\theta)\ra_{\theta} = P,
\end{align}
and
\begin{align}
\label{opt_obj_mc}
\mbox{maximize}~ \int_{0}^{2 \pi} 
\left( - \int_{\theta}^{\theta + \psi_{\max }} Z_1(\phi) d\phi \right) u(\theta) d\theta, ~ \mbox{s.t.}~\la u^2 (\theta)\ra_{\theta} = P,
\end{align}
respectively.
In Eq.~(\ref{opt_obj_mc}), $\psi_{\max}$ represents the phase point that takes the maximum 
value of the potential $v(\psi)$ defined in Eq.~(\ref{eq:std_prob}) with $\tilde{\Delta}_e = 0$ (see also the Appendix in Ref.~\cite{kato2020semiclassical}). 

In order to analyze these problems in a unified way, we consider a general optimization problem,
\begin{align}
\label{opt_obj_ge}
\mbox{minimize}~ \int_{0}^{2 \pi} g(\theta) u(\theta) d\theta, ~ \mbox{s.t.}~\la u^2 (\theta)\ra_{\theta} &= P,
\end{align}
where $g(\theta) = Z_1(\theta)$,
$g(\theta) = {Z}'_1(\theta) $ 
and $g(\theta) = \int_{\theta}^{\theta + \psi_{\max}} Z_1(\phi) d\phi$
for optimization problems of improving 
maximum locking ranges \cite{harada2010optimal},
entrainment stability \cite{zlotnik2013optimal}, and enhancement of phase coherence,
respectively \cite{pikovsky2015maximizing}.
We use the Fourier series to write $g(\theta)$ as
\begin{align}
g(\theta) &= \frac{g^0_{k}}{2} + 
\sum_{k = 1}^{\infty} ( g^a_{k} \cos k\theta + g^b_{k} \sin k \theta ),
\end{align}
with
\begin{align}
g^a_k &= 
\frac{1}{\pi} \int_{0}^{2\pi}  g(\theta) \cos k \theta d\theta,~ 
g^b_k = \frac{1}{\pi} \int_{0}^{2\pi}  g(\theta) \sin k \theta  d\theta. 
\end{align}
Then, the optimization problem given by Eq.~(\ref{opt_obj_ge}) is reformulated 
as follows:
\begin{align}
&\underset{u^a_0, u^b_0, u^a_1, u^b_1, \cdots}{\mbox{minimize}}\ 
\frac{g^a_0 u^a_0}{4} + \frac{1}{2}\sum_{k = 1}^{\infty} ( g^a_k u^a_k + g^b_k u^b_k )
\quad \mbox{s.t.} \quad
\frac{(u^{a}_0)^2}{4} + \frac{1}{2} \sum_{k = 1}^{\infty} \left( (u^{a}_k)^2 +  (u^{b}_k)^2 \right) = P.
\end{align}
A simple calculation gives $u^a_k \propto g^a_k, u^b_k \propto g^b_k
~(k=0,1,2,\ldots)$ as the solution, resulting in 
\begin{align}
u(\theta) =
\sqrt{\frac{P}{ \la g(\theta) ^2 \ra_\theta }} g(\theta),
\end{align}
which reproduces the previous three results obtained by the calculus of variations~\cite{harada2010optimal, zlotnik2013optimal, pikovsky2015maximizing}.

%

\end{document}